\documentclass[12pt]{article}
\usepackage{wrapfig}
\usepackage{epsfig}
\usepackage{hyperref}
\usepackage{amssymb}
\usepackage{amsmath,mathtools}
\usepackage{amsfonts}
\usepackage{graphics}
\usepackage{latexsym}
\usepackage{multirow}
\usepackage{fixmath}
\usepackage{color}
\usepackage{stackrel}
\usepackage{txfonts}
\usepackage{centernot}
\usepackage{mathabx}
\usepackage{slashed}
\usepackage{bbm}
\usepackage[low-sup]{subdepth}

\newcommand{\bea}{\begin{eqnarray}}
\newcommand{\eea}{\end{eqnarray}}
\newcommand{\bite}{\begin{itemize}}
\newcommand{\eite}{\end{itemize}}

\newcommand{\phibar}{\hskip1.0mm{\bar{\hskip-1.0mm\phi}}}
\newcommand{\phitilde}{\hskip1.0mm{\tilde{\hskip-1.0mm\phi}}}
\newcommand{\phitildesub}{\hskip0.5mm{\tilde{\hskip-0.5mm\phi}}}
\newcommand{\MSbar}{\,\overline{\!\rm MS\!}\;}

\date{}

\begin{document}
\title{
\vspace{-1.5cm} 
\flushleft
    \phantom{\normalsize DESY-23-101} \\
\vspace{0.5cm}
\centering{\bf Repercussions of the Peccei-Quinn axion on QCD}}

\author{G.~Schierholz\\[1em] 
Deutsches Elektronen-Synchrotron DESY,\\ Notkestr. 85, 22607 Hamburg, Germany}

\maketitle
\vspace*{-0.5cm}
\begin{abstract}
  The axion, originally postulated by Peccei and Quinn to solve the strong CP problem, has become of great interest in particle and astroparticle phenomenology. Yet it has a problem. It is widely assumed that the axion leaves the nonperturbative features of QCD unscathed. This is, however, not the case. On the contrary, the axion extension is found to be in conflict with seminal results on the low-energy properties of QCD. The key is that the anomalous coupling of the axion to the gauge bosons can be integrated out, leaving behind a path integral over topologically trivial gauge potentials.
\end{abstract}
\vspace*{0.5cm}

\section{Introduction}

Quantum chromodynamics (QCD) decribes the strong interactions remarkably well, from the shortest distances probed so far to macroscopic distances shaped by the topological properties of the theory. Yet it faces a problem. The theory allows for a CP-violating term $S_\theta$ in the action,
\begin{equation}
  S = S_{\rm QCD} + S_\theta \,. 
  \label{S1}
\end{equation}
In Euclidean space-time it reads
\begin{equation}
S_\theta = - i\, \theta\, Q\,, \quad Q = \int d^4x\, P\, \in\, \mathbb{Z}\,, 
\label{S2}
\end{equation}
where $Q$ is the topological charge and 
\begin{equation}
 P=-\frac{1}{32\pi^2}\, \epsilon_{\mu\nu\rho\sigma} \textrm{Tr}\left[F_{\mu\nu} F_{\rho\sigma}\right]\,,
\label{P}
\end{equation}
is the Pontryagin density. In this formulation $\theta$ enters as an arbitrary phase, referred to as the vacuum angle, which can take values $-\pi < \theta \leq +\pi$. Thus, there is the possibility of CP violation in the strong interaction. A nonvanishing value of $\theta$ would result in an electric dipole moment $d_n$ of the neutron. To date the most sensitive measurements of $d_n$ are compatible with zero. The current upper bound is $|d_n| < 1.8 \times 10^{-13} e\,\textrm{fm}$~\cite{Abel:2020pzs}, indicating that $\theta$ is anomalously small at hadronic scales. This puzzle is referred to as the strong CP problem, which is one of the most intriguing unsolved problems in the elementary particles field.

A popular view is that the solution of the strong CP problem lies beyond QCD and the Standard Model. The first instinct in such a situation is to propose a new symmetry that suppresses, or rules out, CP-violating processes in the strong interactions. Indeed, Peccei and Quinn~\cite{Peccei:1977hh} concocted a global chiral U(1) symmetry, which effectively rotates the $\theta$ parameter into a dynamical, CP-conserving field $\phi$, the axion. This is realized by augmenting the $\theta$ term (\ref{S2}) with the axion action~\cite{Peccei:2006as}, 
\begin{equation}
  S_\theta \rightarrow S_\theta + S_\phi = \int d^4x \, \left[\frac{1}{2} \big(\partial_\mu \phi\big)^2 - i \left(\theta + \frac{\phi}{f}\right) \, P + L_\phi(\partial_\mu \phi\, J_\mu^5)\right]\,,
    \label{PQ}
\end{equation}
where $f$ is a scale factor, and the Lagrangian $L_\phi$ contains the derivative coupling to the light quark axial vector currents, collectively denoted by $J_\mu^5$. We have put all CP violation in $S_\theta$ and deal with real quark mass matrices only. To leading order in $1/f$ the interaction with the $u$ and $d$ quarks is of the form~\cite{Weinberg:1996kr}
\begin{equation}
  L_\phi = \frac{f_u}{f}\, \partial_\mu\phi\, \bar{u}\gamma_5\gamma_\mu u + \frac{f_d}{f}\, \partial_\mu\phi\, \bar{d}\gamma_5\gamma_\mu d \,,
  \label{W}
\end{equation}
where $f_u$ and $f_d$ are dimensionless coupling constants. The axion decay constant is estimated to be $f = 10^{11} - 10^{18} \, \textrm{GeV}$~\cite{Kawasaki:2018qwp}. For the most general form of the interaction respecting the U(1) symmetry see, for example,~\cite{Kim:2008hd}. It is readily seen that the path integral is invariant under the global change of integration variables $\phi(x) \rightarrow \phi(x)+\delta$, called shift symmetry, which practically eliminates $\theta$ from the action. Beyond the original idea, the axion has aroused great interest in particle and astroparticle phenomenology. For example, the axion is considered a candidate for dark matter. Axions also arise in fundamental theories such as string theory. This has prompted laboratory, astrophysical and cosmological searches for axions and axion-like particles~\cite{Graham:2015ouw}.

In all this effort it has been assumed that the axion extension of the Standard Model does not change the nonperturbative properties of the theory. That can be doubted. There is strong evidence, both from purely theoretical considerations and numerical investigations~\cite{Callan:1979bg,tHooft:1981bkw,Cardy:1981qy,Nakamura:2021meh}, for a nontrivial $\theta$ dependence, which is incompatible with the shift symmetry. Already in 1979 't Hooft~\cite{tHooft:1981bkw} predicted at least one phase transition at $|\theta| < \pi$. This suggests that the axion will seriously affect the topological properties of QCD, which questions the consistency of the model.

In the following we treat the Peccei-Quinn theory (\ref{PQ}) as an effective theory valid at QCD scales. It can be assumed to be decoupled from the Yukawa couplings to the Higgs fields by the Appelquist-Carazzone decoupling theorem~\cite{Appelquist:1974tg}. This makes studying the effects of the axion field on the topological properties of QCD a well defined task.


\section{Path integral}

We may drop the $\theta$ term in (\ref{PQ}) now. This leaves us with the action
\begin{equation}
  S = \int d^4x \, \left[\frac{1}{2}\, \big(\partial_\mu \phi\big)^2 - \frac{i}{f}\, \phi\, P + L_\phi(\partial_\mu \phi\, J_\mu^5)\right] + S_{\rm QCD}\,.
  \label{action}
\end{equation}
The lattice regularization of the path integral provides a nonperturbative definition of the theory, free of infrared and ultraviolet singularities, which enables direct and mathematically well defined calculations. In the following we implicitly assume that space-time is replaced by a four-dimensional lattice of volume $V$, though we use continuum notation. We imagine the lattice spacing to be small enough, so that tunneling between different topological sectors is excluded, like in the continuum. The key ingredient is the generating functional 
\begin{equation}
  \int\! \mathcal{D}[A_\mu] \int\! \mathcal{D}[\phi] \, \int\, [\cdots] \, \exp\Big\{\!-\!\int d^4x \,\Big[\frac{1}{2} \big(\partial_\mu \phi\big)^2 - \frac{i}{f}\, \phi\, P+ L_\phi(\partial_\mu \phi\, J_\mu^5)\Big] - S_{\rm QCD}\Big\}\,,
  \label{gf}
\end{equation}
where \raisebox{1.75pt}{$\,\int {\scriptstyle [\cdots]}$} stands for the integral over the fermion fields. The fermions need to be integrated out first, resulting in the fermion determinant, before the integrals over $\phi$ and the gauge fields can actually be performed. Physical quantities, like Green's functions, will not depend on the mean value of $\phi$. 
Writing
\begin{equation}
    \phi(x) = \phibar + \phitilde(x)\,, \quad \phibar \equiv \langle \phi\rangle =\frac{1}{V} \int d^4x\, \phi(x) \,,
    \label{cov}
\end{equation}
we may divide the path integral over the axion field $\phi(x)$ into an integral over $\phibar$ and a path integral over the field $\phitilde(x)$ orthogonal to $\phibar$,
\begin{equation}
 \int \mathcal{D}[\phi]\, \cdots\, = \int_{\langle \phitildesub \rangle=0} \!\!\! \mathcal{D}[\phitilde] \int d\phibar\, \cdots \,.
\end{equation}
Accordingly, the anomalous contribution to the action (\ref{PQ}) divides into two pieces,
\begin{equation}
  \int d^4x \, \phi(x)\, P(x) = \phibar\, Q + \int d^4x \, \phitilde(x)\, P(x)\,.
  \label{split}
\end{equation}
The kinetic term and the derivative couplings do not depend on $\phibar$. We thus can carry out the integral over $\phibar$. Due to the periodicity of the integrand, the integration can be restricted to the finite interval $[-f \pi,f \pi]$. It results in a Kronecker delta function, $2\pi f\delta_{Q\, 0}$, which restricts the gauge fields to the sector of zero topological charge, $Q=0$. Up to a factor, we end up with the generating functional 
\begin{equation}
  \int_{Q=0}\! \mathcal{D}[A_\mu] \int_{\langle \phi\rangle=0}\! \mathcal{D}[\phi] \, \int\, [\cdots] \, \exp\Big\{\!-\!\int d^4x \,\Big[\frac{1}{2} \big(\partial_\mu \phi\big)^2 - \frac{i}{f}\, \phi\, P+ L_\phi(\partial_\mu \phi\, J_\mu^5)\Big] - S_{\rm QCD}\Big\}\,,
  \label{gf2}
\end{equation}
where we have changed $\phitilde$ to $\phi$. Restricting the integral to $\langle \phi\rangle =0$ is common practice in the calculation of the constraint effective potential~\cite{ORaifeartaigh:1986axd}. Continuum-like gauge fields split naturally into quantum mechanically disconnected sectors of fixed topological charge. On the lattice that happens at lattice spacings $a \lesssim 0.05\, \textrm{fm}$. Hence, the limitation to trivial topology is just a matter of picking the sector $Q=0$. The path integral over $\phi$ in (\ref{gf2}) can be sampled by a Gaussian white noise, which is an ergodic random process.

One might argue that integrating the gauge fields and the white noise first will change the result. This is not the case. The point is that the Lagrangian $L_\phi$ does not depend on $\phibar$, as the axion does not couple to the zero modes~\cite{Gelmini:1982zz}, and no such interaction can be generated. The path integral over the gauge fields splits into disconnected sectors of charge $Q$, which results in the partition function
\begin{equation}
  Z = \int d\phibar\, \sum_Q \exp\left\{\frac{i}{f} \phibar\, Q\right\} P(Q) \equiv \int d\phibar\, P(\phibar)\,.
  \label{Z}
\end{equation}
$P(Q)$ is the probability for charge $Q$ (at $\phibar=0$). It includes the interaction of the axion (here the white noise) with the quarks, in addition to the usual QCD interactions. The pseudo-distribution $P(\phi)$ defines the (constraint) effective potential~\cite{Fukuda:1974ey,ORaifeartaigh:1986axd}
\begin{equation}
  V U_{\rm eff}(\phibar) = - \log P(\phibar) + c \,.
\end{equation}
It is entirely determined by $P(Q)$ up to a constant. Assuming a Gaussian distribution for $P(Q)$, $P(Q)=(1/\sqrt{2\pi\langle Q^2\rangle}) \exp\{-Q^2/2\langle Q^2\rangle\}$, this gives the common result
\begin{equation}
  U_{\rm eff}(\phibar)= \frac{1}{2f^2}\, \chi_t\, \phibar^2 \,,
  \label{eff}
\end{equation}
where $\chi_t=\langle Q^2 \rangle/V$ is the topological susceptibility. It should be noted that $P(Q)$ ceases to be a Gaussian in the larger volume~\cite{Ce:2015qha,Nakamura:2021meh}, not taking axion interactions into account. We are interested in the effect on QCD observables. The expectation value of an observable $\mathcal{O}$ is
\begin{equation}
  \langle \mathcal{O} \rangle = \frac{1}{Z} \int d\phibar \sum_Q\, \exp\left\{\frac{i}{f} \phibar\, Q\right\} \langle \mathcal{O}\rangle_Q\,P(Q)\,,
  \label{exp}
\end{equation}
where $\langle \mathcal{O}\rangle_Q$ is the value in the sector of charge $Q$. On the finite lattice one can interchange integration and summation. This gives $\langle \mathcal{O} \rangle = \langle \mathcal{O} \rangle_{Q=0}$, which is identical with integrating $\phibar$ first. 

We can go one step further now and cast the remainder of the anomalous coupling $\phi P$ ($\phi = \phitilde$ here) in (\ref{gf2}) into the Lagrangian $L_\phi$ by partial integration.
The Pontryagin density is known to be a total divergence, $P=\partial_\mu \Omega_\mu^{\scriptscriptstyle (0)}$, where
\begin{equation}
  \Omega_\mu^{\scriptscriptstyle (0)} = -\frac{1}{8\pi^2}\, \epsilon_{\mu\nu\rho\sigma}\, \text{Tr} \left[A_\nu \partial_\rho A_\sigma + \frac{2}{3} A_\nu A_\rho A_\sigma\right]
\end{equation}
is the Chern-Simons density or $0$-cochain~\cite{Bardeen:1974ry,Laursen:1985cn}. For purposes of the topological analysis we compactify space-time. We consider  the 4D torus $T$~\cite{Luscher:1981zq,Gockeler:1985dp,Kronfeld:1986ts}, which is defined by $T = \{x\,|\,0\leq x_\mu \leq L_\mu\,;\, \mu=1, \cdots , 4\}$ with periods $L_\mu$. We impose the condition that the gauge fields be regular everywhere, except at the boundaries. By partial integration we thus get
\begin{equation}
  \int_{T} d^4x\, \phi\, P = \int_{\partial T} d^3\sigma_\mu\, \phi\, \Omega_\mu^{\scriptscriptstyle (0)} - \int_{T} d^4x\, \partial_\mu \phi \, \Omega_\mu^{\scriptscriptstyle (0)}\,,
  \label{part1}
\end{equation}  
where $\partial T$ denotes the boundary of $T$. On the torus, periodicity of the action implies periodic boundary conditions for the axion and periodic boundary conditions up to a gauge transformation $\varv_\mu=g(x_\mu=L_\mu)\,g^{-1}(x_\mu=0)$ for the gauge fields. The gauge transforms $\varv_\mu$ can be categorized into homotopy classes labelled by the topological charge $Q \in \pi_3(\textrm{SU(3)})=\mathbb{Z}$. For $Q=0$ the transformation $\varv_\mu$ is homotopically trivial, which means that it can be continuously deformed to $\varv_\mu=1$~\cite{Jackiw:1976pf}. The result is $\int_{\partial T}\! d^3\sigma_\mu\, \phi\, \Omega_\mu^{\scriptscriptstyle (0)} = 0$. The second term on the right-hand side of (\ref{part1}) can be absorbed in the Lagrangian,
\begin{equation}
  L_\phi + \frac{i}{f} \partial_\mu \phi\, \Omega_\mu^{\scriptscriptstyle (0)} \rightarrow  L_\phi\,,
\end{equation}
replacing $J_\mu^5 \rightarrow J_\mu^5 + (i/f)\, \Omega_\mu^{\scriptscriptstyle (0)}$. This leads us to our final result for the generating functional 
\begin{equation}
  \int_{Q=0}\! \mathcal{D}[A_\mu] \int\! \mathcal{D}[\phi] \, \int\, [\cdots] \, \exp\Big\{\!-\!\int d^4x \,\Big[\frac{1}{2} \big(\partial_\mu \phi\big)^2 \! + L_\phi(\partial_\mu \phi\, J_\mu^5)\Big] - S_{\rm QCD}\Big\}\,,
  \label{gf3}
\end{equation}
where we can relax the constraint $\langle \phi \rangle = 0$ now. It should be noted though that in the numerical analysis special measures have to be taken to keep $\langle \phi \rangle$ from running to infinity. A similar problem arises in noncompact lattice QED~\cite{Gockeler:1989wj,Horsley:2015eaa}.

The Chern-Simons density $\Omega_\mu^{(0)}$ is a gauge variant quantity, and one might wonder if the action is gauge invariant. Under a gauge transformation $A_\mu \rightarrow g^{-1} (A_\mu + \partial_\mu) g$~\cite{Laursen:1985cn}
\begin{equation}
 \Omega_\mu^{\scriptscriptstyle (0)} \rightarrow \Omega_\mu^{\scriptscriptstyle (0)} - \frac{1}{8\pi^2}\, \varepsilon_{\mu\nu\rho\sigma} \, \partial_\nu \text{Tr} \left[\partial_\rho g\, g^{-1} A_\sigma\right] - \frac{1}{24\pi^2}\, \varepsilon_{\mu\nu\rho\sigma} \text{Tr} \left[\partial_\nu g \,g^{-1} \partial_\rho g \,g^{-1} \partial_\sigma g \,g^{-1}\right]\,.
  \label{gauge}
\end{equation}
The last two terms can be written as a total divergence,  $\Omega_\mu^{\scriptscriptstyle (0)} \rightarrow \Omega_\mu^{\scriptscriptstyle (0)} + \partial_\nu \Omega_{\mu\nu}^{\scriptscriptstyle (1)}$, where $\Omega_{\mu\nu}^{\scriptscriptstyle (1)}$ is the $1$-cochain~\cite{Laursen:1985cn}. This changes the action by the amount $- \int_T d^4x\, \partial_\mu \phi \, \partial_\nu \Omega_{\mu\nu}^{\scriptscriptstyle (1)}$. Partial integration gives
\begin{equation}
  - \int_{\partial T} d^3\sigma_\nu\, \partial_\mu \phi \, \Omega_{\mu\nu}^{\scriptscriptstyle (1)} + \int_T d^4x\, \partial_\mu \partial_\nu\phi \, \Omega_{\mu\nu}^{\scriptscriptstyle (1)}\,.
\end{equation}
The first term vanishes for the same reason $\int_{\partial T} d^3\sigma_\mu\,  \phi \, \Omega_\mu^{\scriptscriptstyle (0)}$ vanishes, the second term vanishes by symmetry, $\Omega_{\mu\nu}^{\scriptscriptstyle (1)}=-\Omega_{\nu\mu}^{\scriptscriptstyle (1)}$.

The limitation to field configurations with $Q = 0$ is largely equivalent to integrating the full QCD partition function over $\theta$,
\begin{equation}
  Z = \frac{1}{2\pi} \int_{-\pi}^{\pi} d\theta\, Z(\theta)\,,
\end{equation}
thus averaging over all $\theta$ vacua, confining and not confining~\cite{Nakamura:2021meh}.

\section{Implications for QCD}

Lattice QCD successfully describes the static, nonperturbative properties of the strong interactions. Many of the key results derive from the nontrivial topology of the gauge fields. The calculations are limited to a finite volume. 
Effective field theory often offers the possibility of continuing the lattice results to the thermodynamic limit. We will show now that the axion extension of QCD is in conflict with well established low-energy properties of the theory.

We focus on the consequences of Eq.~(\ref{gf2}), namely that the path integral of the extended theory projects onto gauge fields with (global) topological charge $Q = 0$, resulting in $\langle \mathcal{O} \rangle = \langle \mathcal{O} \rangle_{Q=0}$ (Eq.~(\ref{exp})). The index theorem connects $Q$ with the zero modes of the Dirac operator $\slashed{D}$,
\begin{equation}
  Q = - \sum_i \int d^4x \, u_i^\dagger(x) \gamma_5 u_i(x) = n_- - n_+\,,
\end{equation}
where $\{u_i(x)\}$ are the zero-mode eigenvectors, 
\begin{equation}
  i\slashed{D}\, u_i(x) = 0\,, \quad \int d^4x \, u_i^\dagger(x) u_i(x) = 1\,,
\end{equation}
of which $n_+$ eigenvectors have positive chirality, $\gamma_5 u_i(x) =  u_i(x)$ and $n_-$ have negative chirality $\gamma_5 u_i(x) = - u_i(x)$. On the lattice $i\slashed{D}$ is realized by the overlap Dirac operator~\cite{Neuberger:1997fp}. The zero-mode eigenfunctions are found to be strongly correlated with instantons of the appropriate charge~\cite{DeGrand:2000gq}. Expressing the topological charge in terms of its zero modes is particularly revealing. A gauge field configuration of positive (negative) topological charge $Q$ has exactly $n_-$ ($n_+$) zero modes. In mathematical language $\textrm{dim\, ker}\, i \slashed{D} = |n_+ - n_-|$. This so-called `vanishing theorem' has been proven in two dimensions~\cite{Ansourian:1977qe,Nielsen:1977aw}, for self-dual configurations in four dimensions~\cite{Brown:1977bj,Carlitz:1978xu}, and for generic metrics and (gauge) connections in two and four dimensions~\cite{Maier}. Moreover, it is an integral part of the chiral effective theory~\cite{Leutwyler:1992yt} with $Z = \sum_Q Z_Q$, an ansatz that is also clearly favored dynamically~\cite{Schierholz:2024var}. The `vanishing theorem' has been confirmed on the lattice in two dimensions~\cite{Chiu:1998bh}, and in four dimensions at zero~\cite{Ilgenfritz:2007xu,Chiu:2011dz,DiGiacomo:2015eva} and nonzero temperature~\cite{Chen:2022fid,Chiu}. It means that any gauge field configuration of charge $Q = 0$ has absolutely no zero modes, which is equivalent to carrying no isolated instantons. 

The projection to $Q = 0$ has far reaching consequences for the Peccei-Quinn extension of QCD. First and foremost
\begin{equation}
  \chi_t = \frac{\langle Q^2\rangle}{V} = 0\,,
  \label{chi}
\end{equation}
a result that holds on any size volume. It holds on any subvolume $V_{\textrm{sub}} \subset V$ as well, as a consequence of the `vanishing theorem'~\cite{Schierholz:2024var}. The topological susceptibility (\ref{chi}) cannot be measured directly, but it enters into several observables. Most significant is the Witten-Veneziano relation~\cite{Witten:1979vv,Veneziano:1979ec}
\begin{equation}
  m_{\eta^\prime}^2 + m_{\eta}^2 - 2 m_K^2 = \frac{12}{f_\pi^2}\, \chi_t \,,
  \label{VW}
\end{equation}
where $\chi_t$ is to be taken in the limit of heavy quarks. A recent result from lattice QCD is $\chi_t = (180.5(5)(43)\, \textrm{MeV})^4$~\cite{Ce:2015qha}, which is in perfect agreement with the left-hand side of (\ref{VW}). Like no other formula, the Witten-Veneziano relation illustrates the role of topological charge in low-energy hadron physics. A quantity of great interest in axion phenomenology is the axion mass. It is given by~\cite{Peccei:2006as}
\begin{equation}
  m_a=\frac{\sqrt{\chi_t}}{f} \,.
  \label{axm}
\end{equation}
Axions are considered a candidate for dark matter. To estimate its mass in the post-inﬂation phase of the Universe, the topological susceptibility needs to be extrapolated to $\textrm{GeV}$ temperatures. This has been possible in finite temperature QCD~\cite{Bonati:2015vqz,Borsanyi:2016ksw}. Assuming that the dark matter is made up of axions, and adopting the lattice QCD result for $\chi_t$, a mass of
\begin{equation}
  m_a=50(4)\, \mu\textrm{eV}
  \label{ma}
\end{equation}
has been suggested~\cite{Borsanyi:2016ksw}. Both results, (\ref{VW}) and (\ref{ma}), are invalidated by the axion dictate~(\ref{chi}).

The absence of the topological charge has profound consequences for the fermion sector. The QCD action has a chiral $\textrm{U}_A(1)$ symmetry,
\begin{equation}
  q \rightarrow \exp\{i \delta \gamma_5\}\,q\,, \quad \bar{q} \rightarrow \bar{q}\, \exp\{i \delta \gamma_5\} \,,
  \label{U1}
\end{equation}
for small $q = u$ and $d$ quark masses that is not realized in the real world. The violation of this symmetry can be attributed to the anomaly in the regularized theory, which has played an important role in establishing QCD as the theory of the strong interactions. In the path integral the anomaly is due to the noninvariance of the measure. Under the change of variables (\ref{U1}), the measure transforms as
\begin{equation}
  \int_{Q=0}\!\!\! \mathcal{D}[A_\mu]   \int\! \mathcal{D}[q] \mathcal{D}[\bar{q}] \rightarrow \int_{Q=0}\!\!\! \mathcal{D}[A_\mu]\, \exp\{i 2 \delta Q\}  \int\! \mathcal{D}[q] \mathcal{D}[\bar{q}] = \int_{Q=0}\!\!\! \mathcal{D}[A_\mu]  \int\! \mathcal{D}[q] \mathcal{D}[\bar{q}] \,,
\end{equation}
Hence, in the case of trivial topology the $\textrm{U}_A(1)$ symmetry is left unbroken. However, a nonvanishing chiral condensate $\langle \bar{q}q \rangle$ does not only break the $\textrm{SU(2)}_L \times \textrm{SU(2)}_R$ symmetry spontaneously but also the axial $\textrm{U}_A(1)$ symmetry.
A spontaneously broken $\textrm{U}_A(1)$ symmetry would require an isoscalar Goldstone boson, the $\eta^\prime$, with a mass~\cite{Weinberg:1996kr}
\begin{equation}
  m_{\eta^\prime} \leq \sqrt{3} \, m_\pi\,.
\end{equation}
Its physical mass, however, is much heavier than the mass of the $\eta$ meson. The puzzle was first solved by 't Hooft~\cite{tHooft:1986ooh}, who showed that by incorporating the newly discovered instantons in the path integral the symmetry gets broken dynamically. The role of the $\textrm{U}_A(1)$ anomaly in QCD phenomenology, and what the loss of it would imply, has been summed up in~\cite{Shore:2007yn}, including the Gell-Mann--Oakes--Renner relation, meson decay constants, the Goldberger-Treiman relation, polarized structure functions and the proton spin crisis. 
The question remains whether chiral symmetry is broken in topologically trivial configurations. Callan, Dashen and Gross~\cite{Callan:1977gz} suggested that chiral symmetry breaking is triggered by the anomaly, which would give 
\begin{equation}
  2 m_q \langle \bar{q} q \rangle = \frac{\langle Q^2 \rangle}{V}
  \label{cc}
\end{equation}
for two light flavors. In the finite volume this relation has been verified by lattice simulations~\cite{Bruno:2014ova}. Further support comes from a study of the finite temperature chiral phase transitions~\cite{Aoki:2021qws}, which showed that both the connected and disconnected chiral susceptibilities are largely determined by the anomaly. If not completely, we would expect the quark condensate to be drastically reduced in the Peccei-Quinn theory, as far as the lattice can tell us. 

\begin{figure}[t!]
  \vspace*{-0.5cm}
  \begin{center}
    \hspace*{-0.5cm}\epsfig{file=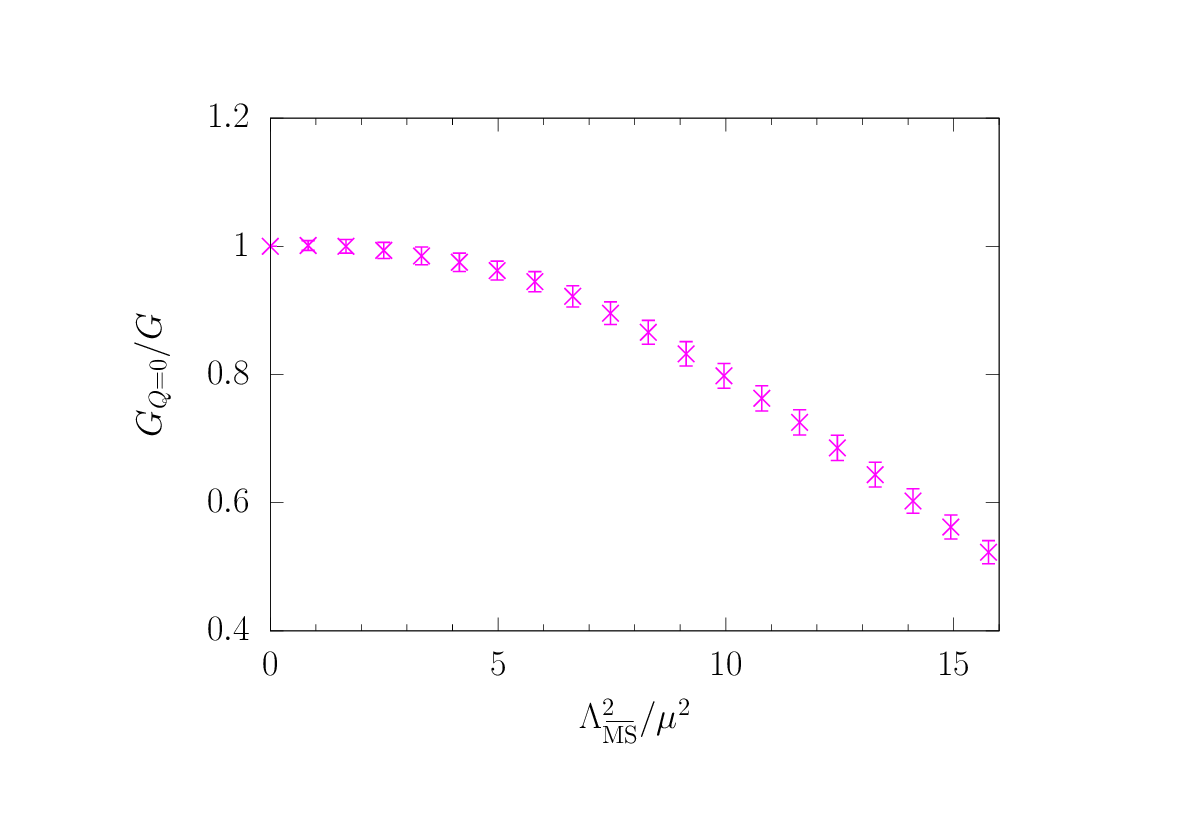,width=9.25cm,clip=}\hspace*{-1.25cm}
    \epsfig{file=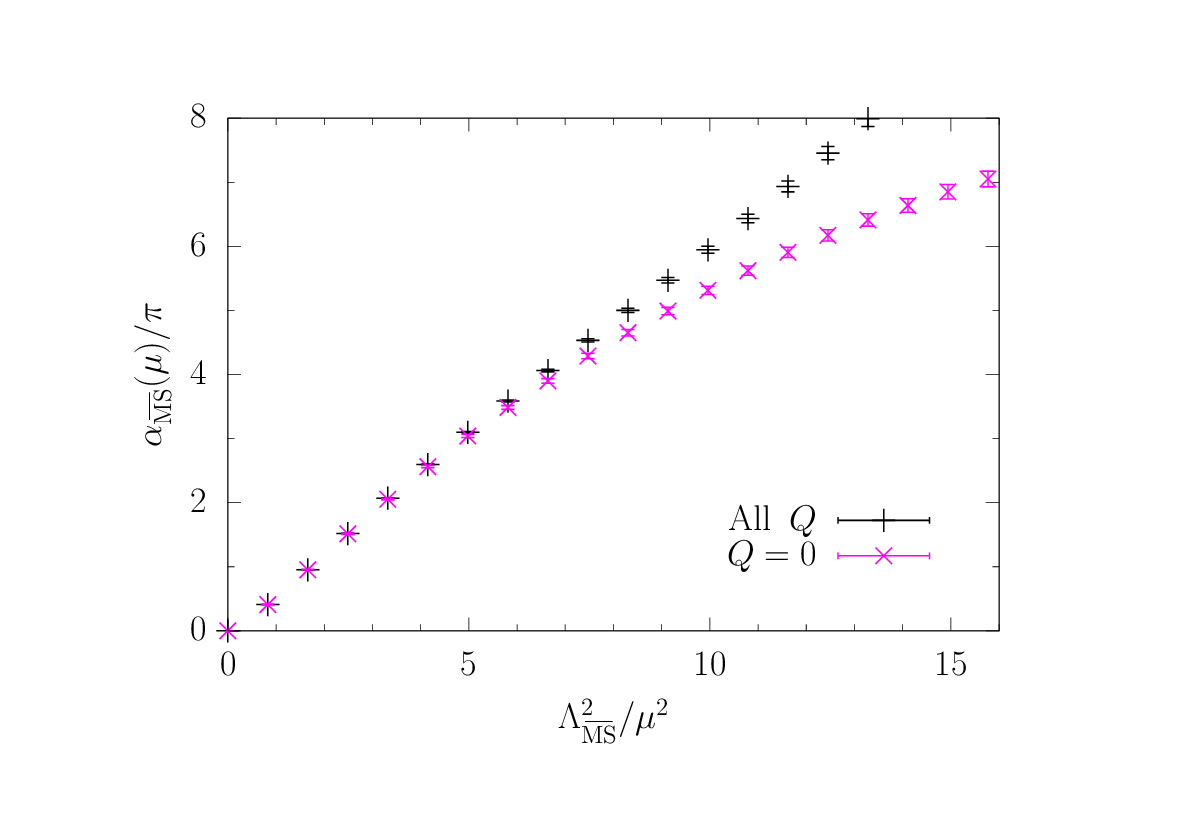,width=9.25cm,clip=}
  \end{center}
  \vspace*{-1.0cm}
  \caption{Left panel: The ratio of gluon condensates for zero and all charges, $G_{Q=0}/G$, as a function of the scale parameter $\mu$. Right panel: The running coupling constant in the $\MSbar$ scheme separately for zero ($\textcolor{magenta}{\times}$) and all charges ($+$), as a function of the scale parameter $\mu$ as well. The data are taken from~\cite{Nakamura:2021meh} on the $32^4$ lattice with lattice spacing $a=0.082 \, \textrm{fm}$.}
  \label{fig}
\end{figure}

There are indications that the Peccei-Quinn theory does not even confine. Consider the pure SU(3) Yang-Mills theory. Color confinement can be attributed to the increase of the strong coupling constant beyond bounds at large distances. This is sometimes referred to as infrared slavery. The gradient flow allows to follow the effective, running coupling $\alpha_s(\mu)$ to very large flow times $t = 1/8\mu^2$. It has been shown~\cite{Nakamura:2021meh,Schierholz:2024var} that  
\begin{equation}
  \alpha_s(\mu) = \frac{\Lambda^2}{\mu^2}
  \label{alpha}
\end{equation}
for sufficently small values of scale parameter $\mu$. A formal proof of (\ref{alpha}) is given in~\cite{Schierholz:2024var}, under the condition of a nonvanishing gluon condensate
\begin{equation}
  G = \frac{\alpha_s}{\pi}\, \langle G_{\mu\nu}^a G_{\mu\nu}^a\rangle \,.
\end{equation}
The essential point here is that the gluon condensate $G$ is independent of the scale parameter $\mu$. Using (\ref{alpha}), a calculation of the static quark-antiquark potential has led to the string tension $\sqrt{\sigma} = 445(19)\,\textrm{MeV}$~\cite{Nakamura:2021meh}, in perfect agreement with phenomenology, and in line with our view of confinement. This is not true anymore in the theory truncated to $Q=0$. In Fig.~\ref{fig} I show the gluon condensate and the running coupling for $Q=0$ with respect to the full ensemble. The gluon condensate $G_{Q=0}$ turns out to be no longer scale invariant, but goes to zero in the infrared. This does not come as a surprise, as the vacuum turns into a semi-classical ensemble of sign-coherent (anti-)instantons at large flow times~\cite{Nakamura:2021meh}. The effective coupling $\alpha_{s,\,Q=0}(\mu)$ is also drastically reduced at large distances, far from a linear increase. The loss of confinement in topologically trivial gauge potentials is familiar. In the two-dimensional $CP^3$ model, which is in many respects similar to QCD, the string tension of two static charges of magnitude $e$ turns out to be~\cite{Schierholz:1994pb} $\sigma = F(2\pi e)-F(0)$, where $F$ is the free energy, $F=F(\theta)$. Assuming a Gaussian distribution for topological charge, this gives
\begin{equation}
\sigma = \frac{\langle Q^2\rangle}{V} \left[1-\cos(2\pi e)\right] = \frac{\langle Q^2\rangle}{V}\, 2\pi^2 e^2 + O(e^4) \,,
\end{equation}
which will vanish for $Q = 0$.

\section{Conclusions}

We conclude that when the gauge fields are restricted to trivial topology, $Q \equiv 0$, one obtains a theory that is different from QCD. This has been indirectly confirmed by lattice calculations, which link several of the seminal features of the theory to the nontrivial topology of the gauge fields. The impact of the $\textrm{U}_A(1)$ anomaly on other low-energy properties of QCD has been extensively reviewed in the literature, for example in~\cite{Shore:2007yn}. Thus, when extending the Standard Model care must be taken to ensure that the basic values of the theory are not changed.

A lattice simulation of the model would be desirable. So far this is not possible though, due to the complex nature of the action. In~\cite{Nakamura:2018pxj} we have performed a calculation for imaginary couplings $1/f$, which renders the action real, thus making it suitable for Monte Carlo simulations. However, in this case the charge is not integrated to $Q=0$, but to $Q \rightarrow \pm \infty$, which we did not realize at the time. 

\section*{Acknowledgment}

I like to thank Meinulf G\"ockeler for carefully reading the manuscript and for constructive criticism.


\begin{thebibliography}{99}

\bibitem{Abel:2020pzs}
C.~Abel, \textit{et al.}
Phys. Rev. Lett. \textbf{124} (2020) 081803 [arXiv:2001.11966 [hep-ex]].  

\bibitem{Peccei:1977hh}
R.~D.~Peccei and H.~R.~Quinn,
Phys. Rev. Lett. \textbf{38} (1977) 1440.

\bibitem{Peccei:2006as}
R.~D.~Peccei,
Lect. Notes Phys. \textbf{741} (2008) 3
[arXiv:hep-ph/0607268 [hep-ph]].

\bibitem{Weinberg:1996kr}
S.~Weinberg,
{\it The quantum theory of fields. Vol. 2: Modern applications},
Cambridge University Press, 2013.

\bibitem{Kawasaki:2018qwp}
M.~Kawasaki, E.~Sonomoto and T.~T.~Yanagida,
Phys. Lett. B \textbf{782} (2018) 181
[arXiv:1801.07409 [hep-ph]].

\bibitem{Kim:2008hd}
J.~E.~Kim and G.~Carosi,
Rev. Mod. Phys. \textbf{82} (2010) 557
[erratum: Rev. Mod. Phys. \textbf{91} (2019) 049902]
[arXiv:0807.3125 [hep-ph]].

\bibitem{Graham:2015ouw}
P.~W.~Graham, I.~G.~Irastorza, S.~K.~Lamoreaux, A.~Lindner and K.~A.~van Bibber,
Ann. Rev. Nucl. Part. Sci. \textbf{65} (2015) 485
[arXiv:1602.00039 [hep-ex]].

\bibitem{Callan:1979bg}
C.~G.~Callan, R.~F.~Dashen and D.~J.~Gross,
Phys. Rev. D \textbf{20} (1979) 3279.

\bibitem{tHooft:1981bkw}
G.~'t Hooft,
Nucl. Phys. B \textbf{190} (1981) 455.

\bibitem{Cardy:1981qy}
J.~L.~Cardy and E.~Rabinovici,
Nucl. Phys. B \textbf{205} (1982) 1.

\bibitem{Nakamura:2021meh}
Y.~Nakamura and G.~Schierholz,
Nucl. Phys. B \textbf{986} (2023) 116063
[arXiv:2106.11369 [hep-ph]].

\bibitem{Appelquist:1974tg}
T.~Appelquist and J.~Carazzone,
Phys. Rev. D \textbf{11} (1975) 2856.

\bibitem{ORaifeartaigh:1986axd}
L.~O'Raifeartaigh, A.~Wipf and H.~Yoneyama,
Nucl. Phys. B \textbf{271} (1986) 653.

\bibitem{Gelmini:1982zz}
G.~B.~Gelmini, S.~Nussinov and T.~Yanagida,
Nucl. Phys. B \textbf{219} (1983) 31.

\bibitem{Fukuda:1974ey}
R.~Fukuda and E.~Kyriakopoulos,
Nucl. Phys. B \textbf{85} (1975) 354.

\bibitem{Ce:2015qha}
M.~C\`e, C.~Consonni, G.~P.~Engel and L.~Giusti,
Phys. Rev. D \textbf{92} (2015) 074502
[arXiv:1506.06052 [hep-lat]].

\bibitem{Bardeen:1974ry}
W.~A.~Bardeen,
Nucl. Phys. B \textbf{75} (1974) 246.

\bibitem{Laursen:1985cn}
M.~L.~Laursen, G.~Schierholz and U.~J.~Wiese,
Commun. Math. Phys. \textbf{103} (1986) 693.


\bibitem{Luscher:1981zq}
M.~L\"uscher,
Commun. Math. Phys. \textbf{85} (1982) 39.

\bibitem{Gockeler:1985dp}
M.~G\"ockeler, M.~L.~Laursen, G.~Schierholz and U.~J.~Wiese,
Commun. Math. Phys. \textbf{107} (1986) 467.

\bibitem{Kronfeld:1986ts}
A.~S.~Kronfeld, M.~L.~Laursen, G.~Schierholz and U.~J.~Wiese,
Nucl. Phys. B \textbf{292} (1987) 330.

\bibitem{Jackiw:1976pf}
R.~Jackiw and C.~Rebbi,
Phys. Rev. Lett. \textbf{37} (1976) 172.

\bibitem{Gockeler:1989wj}
M.~G\"ockeler, R.~Horsley, E.~Laermann, P.~E.~L.~Rakow, G.~Schierholz, R.~Sommer and U.~J.~Wiese,
Nucl. Phys. B \textbf{334} (1990) 527.

\bibitem{Horsley:2015eaa}
R.~Horsley, Y.~Nakamura, H.~Perlt, D.~Pleiter, P.~E.~L.~Rakow, G.~Schierholz, A.~Schiller, R.~Stokes, H.~St\"uben, R.~D.~Young and J.~M.~Zanotti, 
J. Phys. G \textbf{43} (2016) 10LT02
[arXiv:1508.06401 [hep-lat]].


\bibitem{Neuberger:1997fp}
H.~Neuberger,
Phys. Lett. B \textbf{417} (1998) 141
[arXiv:hep-lat/9707022 [hep-lat]];
Phys. Lett. B \textbf{427} (1998) 353
[arXiv:hep-lat/9801031 [hep-lat]].

\bibitem{DeGrand:2000gq}
T.~A.~DeGrand and A.~Hasenfratz,
Phys. Rev. D \textbf{64} (2001) 034512
[arXiv:hep-lat/0012021 [hep-lat]].

\bibitem{Ansourian:1977qe}
M.~M.~Ansourian,
Phys. Lett. B \textbf{70} (1977) 301.

\bibitem{Nielsen:1977aw}
N.~K.~Nielsen and B.~Schroer,
Nucl. Phys. B \textbf{127} (1977) 493.

\bibitem{Brown:1977bj}
L.~S.~Brown, R.~D.~Carlitz and C.~k.~Lee,
Phys. Rev. D \textbf{16} (1977) 417.

\bibitem{Carlitz:1978xu}
R.~D.~Carlitz and C.~k.~Lee,
Phys. Rev. D \textbf{17} (1978) 3238.

\bibitem{Maier}
S.~Maier,
Commun. Math. Phys. \textbf{188} (1997) 407.

\bibitem{Leutwyler:1992yt}
H.~Leutwyler and A.~V.~Smilga,
Phys. Rev. D \textbf{46} (1992) 5607.

\bibitem{Schierholz:2024var}
G.~Schierholz,
[arXiv:2403.13508 [hep-ph]].

\bibitem{Chiu:1998bh}
T.~W.~Chiu,
Phys. Rev. D \textbf{58} (1998) 074511
[arXiv:hep-lat/9804016 [hep-lat]].

\bibitem{Ilgenfritz:2007xu}
E.~M.~Ilgenfritz, K.~Koller, Y.~Koma, G.~Schierholz, T.~Streuer and V.~Weinberg,
Phys. Rev. D \textbf{76} (2007) 034506
[arXiv:0705.0018 [hep-lat]].

\bibitem{Chiu:2011dz}
T.~W.~Chiu, T.~H.~Hsieh and Y.~Y.~Mao,
Phys. Lett. B \textbf{702} (2011) 131
[arXiv:1105.4414 [hep-lat]]; T.~W.~Chiu,
PoS \textbf{LATTICE2019} (2020) 133
[arXiv:2002.06126 [hep-lat]].

\bibitem{DiGiacomo:2015eva}
A.~Di Giacomo and M.~Hasegawa,
Phys. Rev. D \textbf{91} (2015) 054512
[arXiv:1501.06517 [hep-lat]].

\bibitem{Chen:2022fid}
Y.~C.~Chen, T.~W.~Chiu and T.~H.~Hsieh,
Phys. Rev. D \textbf{106} (2022) 074501
[arXiv:2204.01556 [hep-lat]].

\bibitem{Chiu}
T.~W.~Chiu, private communication.

\bibitem{Witten:1979vv}
E.~Witten,
Nucl. Phys. B \textbf{156} (1979) 269.

\bibitem{Veneziano:1979ec}
G.~Veneziano,
Nucl. Phys. B \textbf{159} (1979) 213.

\bibitem{Bonati:2015vqz}
C.~Bonati, M.~D'Elia, M.~Mariti, G.~Martinelli, M.~Mesiti, F.~Negro, F.~Sanfilippo and G.~Villadoro,
JHEP \textbf{03} (2016) 155
[arXiv:1512.06746 [hep-lat]].

\bibitem{Borsanyi:2016ksw}
S.~Borsanyi, Z.~Fodor, J.~Guenther, K.~H.~Kampert, S.~D.~Katz, T.~Kawanai, T.~G.~Kovacs, S.~W.~Mages, A.~Pasztor, F.~Pittler, J.~Redondo, A.~Ringwald and  K.~K.~Szabo,
Nature \textbf{539} (2016) 69
[arXiv:1606.07494 [hep-lat]].

\bibitem{tHooft:1986ooh}
G.~'t Hooft,
Phys. Rept. \textbf{142} (1986) 357.

\bibitem{Shore:2007yn}
G.~M.~Shore,
{\it The U(1)(A) Anomaly and QCD Phenomenology},
Lect. Notes Phys. \textbf{737} (2008) 235
[arXiv:hep-ph/0701171 [hep-ph]].

\bibitem{Callan:1977gz}
C.~G.~Callan, Jr., R.~F.~Dashen and D.~J.~Gross,
Phys. Rev. D \textbf{17} (1978) 2717.

\bibitem{Bruno:2014ova}
M.~Bruno, S.~Schaefer and R.~Sommer,
JHEP \textbf{08} (2014) 150
[arXiv:1406.5363 [hep-lat]].

\bibitem{Aoki:2021qws}
S.~Aoki, Y.~Aoki, H.~Fukaya, S.~Hashimoto,
C.~Rohrhofer and K.~Suzuki,
PTEP \textbf{2022} (2022) 023B05
[arXiv:2103.05954 [hep-lat]].






\bibitem{Schierholz:1994pb}
G.~Schierholz,
Nucl. Phys. B Proc. Suppl. \textbf{37} (1994) 203
[arXiv:hep-lat/9403012 [hep-lat]].

\bibitem{Nakamura:2018pxj}
Y.~Nakamura and G.~Schierholz,
[arXiv:1802.09339 [hep-lat]].

\end{thebibliography}
\end{document}